\documentclass[sigconf,10pt,letterpaper]{sig-alternate-10pt}
\usepackage{subfigure}
\usepackage{stfloats}
\usepackage{flushend}
\usepackage{booktabs}
\usepackage{blindtext}
\usepackage{scrextend}
\usepackage{filecontents}
\usepackage{geometry}
\usepackage{graphicx}
\usepackage{subfigure}
\usepackage{caption}
\usepackage{amsmath}
\usepackage{amssymb}
\usepackage{cite}
\usepackage{indentfirst}
\usepackage{enumitem}
\usepackage{algorithm}
\usepackage{algorithmic}
\usepackage{multirow}
\usepackage{xcolor}
\usepackage{color}
\usepackage{colortbl}
\usepackage{siunitx}
\usepackage{mathptmx}
\usepackage{authblk}
\usepackage{url}
\usepackage{breakurl}
\usepackage{hyperref}
\hypersetup{hidelinks}

\usepackage{geometry}
\newcommand{\systemname}{{SoilTAG}\xspace}
\definecolor{Gray}{gray}{0.9}

\begin{document}

\title{Detecting Soil Moisture Levels Using Battery-Free Wi-Fi Tag}

\author{Wenli Jiao, Ju Wang, Yelu He, Xiangdong Xi, Xiaojiang Chen}
\affil{Northwest University, Xi'an, China}

\maketitle

\begin{abstract}
Soil sensing plays an important role in increasing agricultural output. Existing soil sensing methods failed to achieve both high accuracy and low cost. This paper presents {\systemname}, a battery-free Wi-Fi tag based high accuracy, low-cost and long range soil moisture sensing system. To measure the soil moisture, our idea is converting changes in soil moisture levels into the frequency response of the tag's resonator. To achieve high accuracy, we optimize key resonator parameters to increase the frequency response sensitivity for sensing even small soil moisture changes. To improve the working range, we design patch antenna array for the tag and employ beamforming and beam nulling at the transceiver side. Experimental results show that \systemname achieves a 2\% high accuracy when the working range is less than 6 m, and when the working range is up to 10 m, it achieves 3.64\% accuracy.
\end{abstract}
\section{Introduction}
Smart agriculture could improve agricultural production efficiency \cite{url1,url2,url3}, increase crop output and reduce resource consumption. Worldwide, only 20\% of agricultural land is irrigated, but this land yields 40\% of the world's crops. Knowing accurate soil moisture can make farmer do timely irrigation and thus improve crop yield. Agriculture is consuming the Earth's available fresh water at a surprising rate: 70\% of blue water from watercourses and groundwater is used for agriculture. The organization reports that three times the level used 50 years ago. It is estimated that irrigation needs for farmers may increase by 19\% in the next thirty years. Actually, a large amount of irrigation water is wasted due to excessive irrigation. Therefore, measuring soil moisture has great significance on saving water. Many agriculture communities recommend farmers to use soil sensors to know accurate soil moisture values for improving crop output and saving irrigation water.

The high cost and inconvenience makes it difficult to measure the soil moisture at a lot of locations in real-world farm field or greenhouse, which limits the adoption of traditional soil sensors. Although some soil sensors are low cost but they have coarse sensing resolution. Recently, several efforts on soil moisture sensing have been made in low cost and convenience by employing Wi-Fi and RFID devices. However, the two solutions have short sensing distance less than two meters and lack scalability for large scale farm field because wireless signals(e.g., Wi-Fi and RFID signal) could go through significant attenuation when passing through soil or backscattered by RFID tags. Thus, if we can expand the sensing distance up to meter-scale without incurring high cost or sacrificing accuracy, it will promote widespread application of soil sensing technology in smart agriculture and thus improve the crop yield and save irrigation water.
\begin{figure}[t]
  \setlength{\abovecaptionskip}{3mm}   
  \setlength{\belowcaptionskip}{-3mm}   
  \centering
  \includegraphics[width=0.75\linewidth]{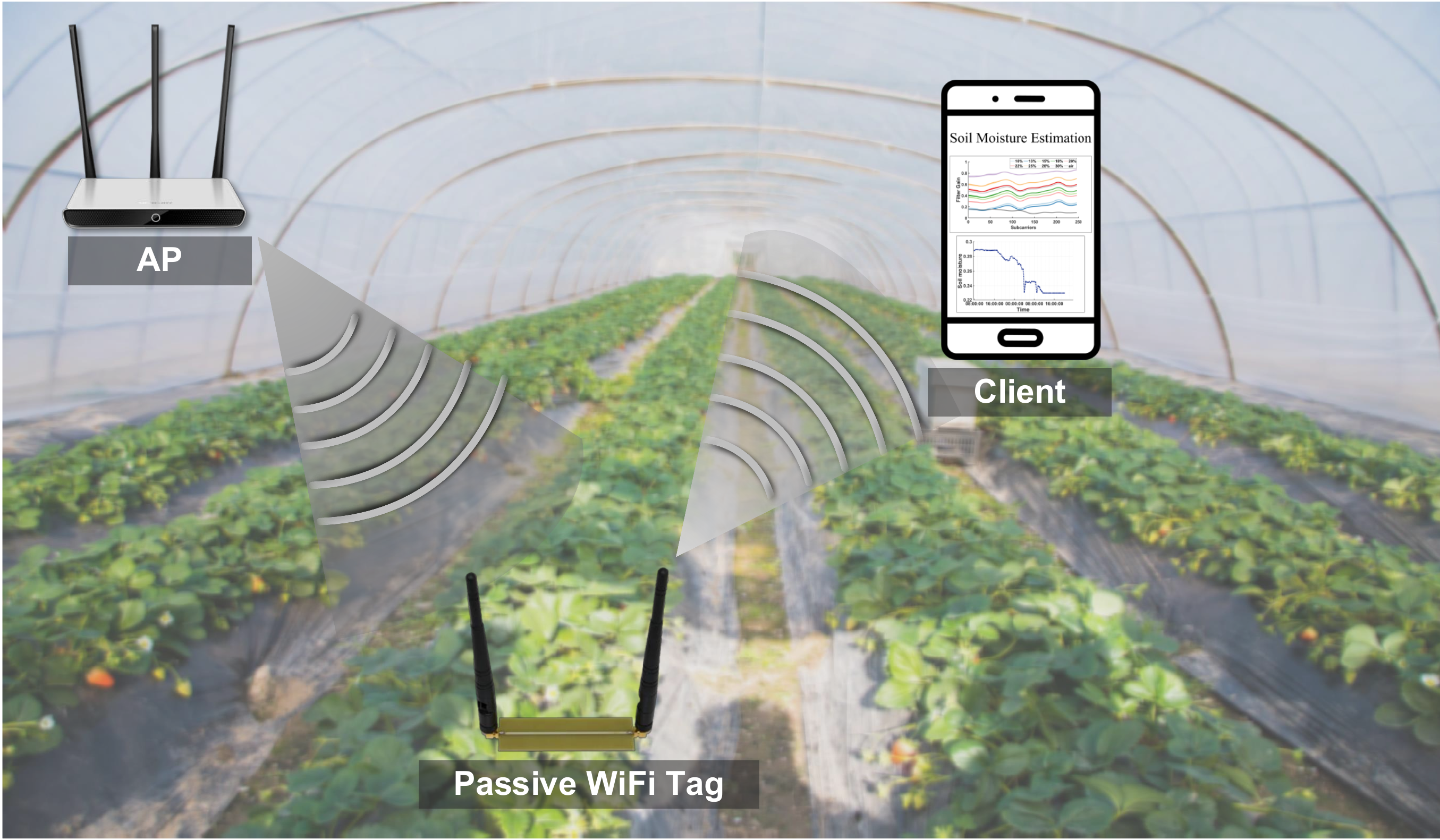}
  \caption{System deployment of {\systemname} in real farm field.}
  \label{fig:app}
\end{figure}

\begin{figure}[t]
\setlength{\abovecaptionskip}{1mm}
\setlength{\belowcaptionskip}{-3mm}
\centering
\includegraphics[width=0.8\linewidth]{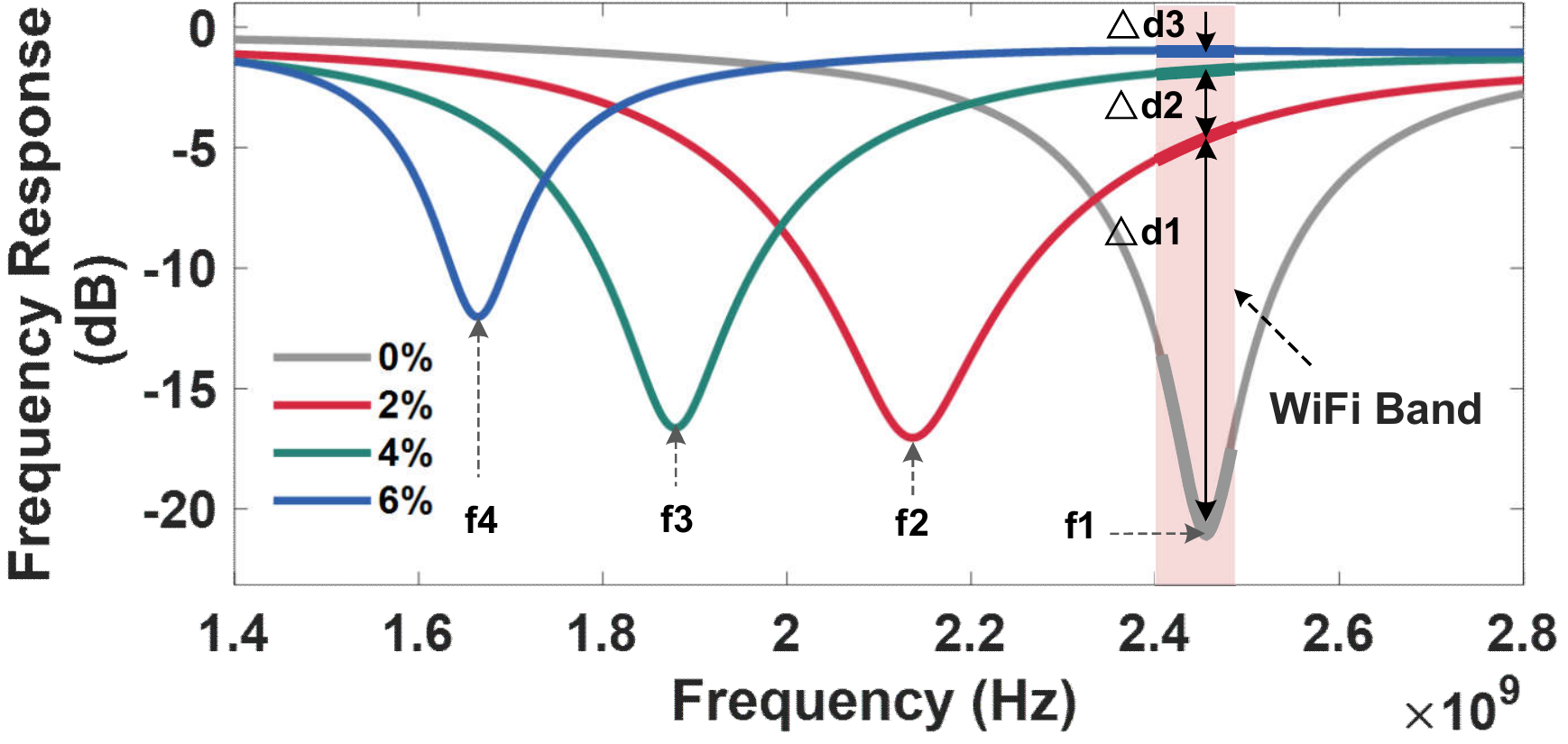}
\caption{Illustration of frequency response features of soil moisture ($0\%, 2\%, 4\%, 6\%$) in different frequency range. Each curve represents one soil moisture level.}
\label{fig:curve}
\end{figure}

Recently, passive sensing technology has been widely adopted for soil moisture sensing~\cite{dey2015novel,gopalakrishnan2020battery} and touch detection \cite{gao2018livetag} due to battery-free and low-cost. The basic idea is that the finger touch or the environmental variation changes the resonant frequency of the passive Wi-Fi tag and thus can success to detect the hand touch through resonance frequency feature. However, due to the available spectrum limitation, existing passive sensing only offer coarse touch sensing. Bringing the passive sensing technology to soil moisture sensing needs about 10 GHz bandwidth. To overcome the bandwidth limitation, our key idea is to convert the resonance frequency characteristics into the amplitude characteristics. Specifically, we identify soil moisture through the amplitude over the Wi-Fi band (i.e., the pink region in Fig.~\ref{fig:curve}). Based on this idea, we design a passive Wi-Fi tag as the soil moisture sensor and propose {\systemname} system -- a passive sensing system that can detect fine-grained soil moisture by identifying the frequency feature of the reflection signal from passive Wi-Fi tags. As shown in Fig.~\ref{fig:app}, the passive tag is attached on the soil or buried in the soil. The Wi-Fi transmitter emits signal to the tag, and the tag reflects it to the Wi-Fi receiver. When the Wi-Fi signal passes through the tag, different soil moisture levels could result in different frequency response feature. By obtaining the channel state information for the reflected signal, the receiver can estimate the soil moisture level.

Even if we can identify soil moisture through the power feature of channel frequency response, we can not detect soil moisture from 0\% to 100\% while ensuring fine-grained soil moisture sensing. This is because that the available bandwidth limits the number of signal features we can observe and therefore limits the measurable soil moisture range. Fig.~\ref{fig:curve} shows the frequency response of the resonator corresponding to $i.e., 0\%, 2\%, 4\%, 6\%$ soil moisture levels. In 1.4--2.8~GHz band, the four curves are distinguishable. However, looking at the Wi-Fi band (i.e., the pink region), we can only distinguish three soil moisture level $0\%, 2\%, 4\%$ and the maximum detectable soil moisture is only $6\%$. This result obviously cannot meet the needs of agriculture for soil moisture measurement.

So, to overcome the limitation of the spectrum bandwidth on the measurable soil moisture range, we propose multi-resonator tag prototype. Specifically, we integrate multiple resonators with different resonant frequencies to one tag that can sense different soil moisture ranges. The key to realize this idea is to optimize the geometric size of resonators. The optimization process is based on our empirical observations - $(i)$ the resonance frequency shifts to low frequencies with the soil permittivity increasing and $(ii)$ the original resonant frequency of resonator must be larger than the Wi-Fi band. In Section, we dive into the details of the optimization process and give two examples of tag design to illustrate our idea. The novel sensor design methodology can be extended to other passive sensor design.

Secondly, in practice the reflection signal of the passive tag suffers from significant LoS and multipath signals interference problem, making it difficult to accurately extract channel frequency response induced by the tag. To overcome this challenge, our basic idea is to boost $Tx-TAG-Rx$ channel and suppressing $Tx-Rx$ and multipath channels. To do so, we resort to beam alignment with the tag.
To enable beam alignment, the difficulty lies in estimating the direction of the chipless tag because the tag is unable to perform channel estimation and feedback its spatial angle to the transmitter. To address this problem, we are inspired by existing work \cite{wei2017Emi}, we leverage both the transmitter and receiver to jointly estimate the spatial angle of the tag. Specifically, the transmitter performs beam scanning while the receiver obtain the beam pattern profile and the multipath profile. By searching the peaks in the two angle spectrum profiles, the receiver can obtain the angle of tag relative to the transmitter and receiver respectively. Based on the obtained angle information, the transmitter and receiver can align beam with the tag. In Sec.~\ref{sec:beam}, we describe the beam alignment algorithm in detail.

We have built a prototype of SoilTAG using WARP software radios and evaluated it empirically with extensive experiments in indoor and outdoor environments respectively. Our experiment result shows that the sensing accuracy can reach $2\%$ when the transmitter-tag distance is less than 4~m.

\textbf{Contributions:} $(i)$ SoilTAG is the first low cost and high accuracy soil moisture sensing system that exploits the interference of the soil to the passive resonator as the fingerprint of soil moisture. By optimizing the sensitivity of the tag and employing machine learning based feature recognition, SoilTAG delivers $2\%$ accuracy using $80$ MHz bandwidth in $2.4$ GHz Wi-Fi band and achieve sensing distance up to 10 meters. $(ii)$ We employ the power feature of channel frequency response to achieve soil moisture sensing without need for large bandwidth. $(iii)$ We build a {\systemname} prototype and conduct extensive experiments to verify its effectiveness in two laboratory rooms and an open field.

\section{Background and Related Work}
\subsection{The relationship between soil moisture and soil permittivity}\label{sec:preliminary}

Current RF-based soil moisture sensing technique (e.g.,TD-R~\cite{ledieu1986method,bhuyan2020soil} and FDR~\cite{linmao2012fdr,bohme2013calibrating}) rely on measuring the soil permittivity because the soil permittivity affects RF signal's transmission property~\cite{ding2019towards,josephson2021low} or the circuit's resonant frequency~\cite{forouzandeh2015chipless}.
The relationship between soil moisture and soil permittivity provide the basis for RF-based soil moisture sensing technique.
In summary, this relationship provides us with a basis for designing a proper sensing tag.
Our work also exploits the affect of soil permittivity on the chipless tag's frequency response to detect different soil moisture levels. 

\textbf{Topp model.} In a water, soil and air mixture, the permittivity is dominated by the amount of water present. Topp model~\cite{topp1980electromagnetic} gives the mathematical relationship as follow: 
\begin{equation}\label{eq:topp}
\epsilon_{soil} = 3.03+9.3*\theta+146*\theta^{2}-76.6*\theta^3.
\end{equation}
where $\varepsilon_{soil}$ is the soil relative permittivity and $\theta$ is the soil moisture. In this paper, \textit{permittivity} refers to the relative permittivity~\cite{permittivity} which is the ratio of the absolute permittivity to the vacuum permittivity.
To accurately measure the soil water content, in this paper, we use the gravimetric water content (GWC\%) as the soil moisture unit. The gravimetric water content is defined as the ratio of the weight of water to the weight of dry soil.
In Sec.~\ref{sec:4.1}, we use the permittivity value computed from Eq.~\eqref{eq:topp} to design the resonator tag.

\subsection{Soil Moisture Measurement}
Existing soil moisture sensing approaches fall into the following three categories:

\textbf{Dedicated soil sensors.}
Currently, commodity soil moisture sensors can be divided into three kinds, including TDR-based \cite{ledieu1986method,bhuyan2020soil}, FDR-based sensors \cite{linmao2012fdr,bohme2013calibrating}, and neutron method based sensor~\cite{heidbuchel2016use,andreasen2017status}. Although they can provide fine-grained soil moisture sensing, they are extremely expensive with a price higher than 50 USD,  which limits the widespread use.
Furthermore, most of them require users to hand-hold, which is very inefficient to measure soil moisture of a large number of locations in a farm field or a greenhouse.  Moreover, the power supply problem of dedicated soil moisture sensors is more challenging in real-world farm filed applications\cite{Cardell2005A}.

To address automatic monitoring, wireless sensor networks were proposed~\cite{barroca2013wireless,Cardell2005A,Norris2008Temperature,Vellidis2008A} where the wireless module is responsible for data transmission with gateways and the sensor is responsible for soil sensing. Although WSNs can achieve large scale soil moisture sensing, sensor nodes require battery power, and a large number of sensor nodes could be very costly. In addition, the soil sensor only provides a coarse-grained soil moisture measurement. 
In contrast, SoilTAG employs low-cost and battery-free passive tags as soil moisture sensors but still provides higher accuracy than low-cost dedicated sensors.

\textbf{RF-based soil moisture sensing.} RF-based soil moisture sensing principle can be split into two main categories: measuring the reflectivity of the radar signal (the reflection coefficient) when impinging the soil surface, or measuring the time of flight (ToF) of the signal when the signal penetrating the soil.
Remote sensing technique based soil moisture sensing method~\cite{mohanty2017soil,wang2009satellite,njoku1996passive} impinges the signal onto the soil surface and measure the reflection coefficient to infer the soil moisture. Although this technique can obtain soil moisture of large scale ground, it is unable to provide a fine-grained spatial resolution. In addition, it can only detect soil moisture on the shallow soil surface with a depth of a few centimeters, and is incapable of detecting the soil moisture at plant roots (30$\sim$75 cm). Hence, remote sensing based soil moisture measurement approach is not suitable for precision agriculture in greenhouse or small farm fields.

To provide finer spatial resolution and reduce the system cost, the recent work Strobe~\cite{ding2019towards} bury the wifi transmitter device into soil and place the receiver on the ground to measure the relative ToF value when Wi-Fi signals pass through the soil.  Although Strobe can achieve relatively accurate soil moisture estimation, it is inconvenient to massively deploy in a real farm field or greenhouse because the antenna array and the receiver radio need to be connected by RF cables. Moreover, the distance between the Wi-Fi transmitter and the receiver is less than 1.5 m, which further limits its real-world application.
Recently, the work~\cite{josephson2021low} proposed a hybrid approach for soil moisture sensing, which employs a commodity radar paired with underground backscatter tags to measure soil moisture. However, this system's measuring range is limited to 30\%, which does not cover the range of soil moisture suitable for the most vegetables (41\%$\sim$80\%)~\cite{moisturerange1}. In addition, the in-ground backscatter tags need battery power, which has the risk of soil contamination from battery corruption.

\textbf{RFID based sensing.} To address the problem of sensor node power supply in practice, past work has focused on backscatter-based RFID technology for battery-free soil moisture sensing~\cite{Aroca2017Application,Aroca2018Calibration,S2018Two}. For example, the GreenTag system captures the DMRT feature of the RFID tag's backscatter signal to measure soil moisture. Although it can measure soil moisture with high accuracy and work well in the greenhouse, it is infeasible for sensing bare soil in a real-world farm field because it relies on sticking RFID tags on the outer wall of the soil pots. Some previous works~\cite{Aroca2017Application,Aroca2018Calibration,S2018Two,wang2020Soil} have studied to attach RFID tags on the soil to measure soil moisture but it can only measure coarse-grained moisture variations.
Unlike existing backscatter-based soil moisture sensing systems, SoilTAG can work under various agricultural scenarios with fewer limitations and also provide high accuracy soil moisture sensing.

\subsection{Chipless Sensing}
Chipless sensing~\cite{Perret2014Radio} has been proposed for sensing soil moisture~\cite{Alonso2017UHF,Fonseca2017A} or air humidity~\cite{Amin2014Development,Nair2013A} by designing chipless UHF tag with special circuit structure. For example, the work~\cite{Alonso2017UHF} explores the feasibility of using a monopole probe to sense soil moisture. Another work~\cite{Fonseca2017A} senses moisture changes by adding a capacitor sensor to an RFID tag. However, these proposals can only sense several relative soil moisture levels rather than an absolute soil moisture value. 
Several literature~\cite{gopalakrishnan2020battery,dey2015novel} leverage resonator-based chipless sensor to sense soil moisture. However, this system requires 670~MHz (i.e., 0.98--1.65 GHz) frequency band for sensing, which has a low spectral efficiency and also rely on Radar devices or the vector network analyzers as readers~\cite{Amin2014Development,Nair2013A}. In addition, they have a short sensing distance. As a result, such a system could be extremely expensive and be infeasible to deploy in practice. Unlike traditional chipless schemes, {\systemname} works in an 80 MHz Wi-Fi band while achieving fine-grained soil moisture sensing. In addition, by employing our sensing design modality, one can adjust the measurable soil moisture range according to the user's need.

Chipless sensing has also been used for sensing touch events. For example, the work LiveTag~\cite{gao2018livetag} designed a chipless Wi-Fi tag to sense a two-state interaction, i.e., touched or not touched. Thus, LiveTag can't sense fine-grained continuous changes. Differs from LiveTag, our SoilTAG system can precisely and continuously measure the soil moisture changes.

\subsection{Beam Alignment}
SoilTAG system is also related to beam alignment technology~\cite{abari2017enabling}. Our beam alignment method for passive tags builds on existing azimuth angle estimation~\cite{wang2013dude} and beamforming technologies~\cite{gao2018livetag}. However, existing feedback-based beam alignment approaches cannot be directly used for our system because passive tags cannot feedback its channel state information. Unlike these applications(eg., MU-MIMO) where the AP can align beams with several target users, we need both transmitter and receiver to direct their beams to a passive tag instead of an active receiver.

SoilTAG is motivated by multipath profile based RFID localization. Unlike RFID localization which utilizes the angle of arriving power profile to estimate the location of a tag, we jointly estimate the angle parameters of the reflection path of the passive tag, including the angle of the tag relative to the transmitter and the receiver respectively. Based on the two angles information, both the Wi-Fi transmitter and receiver can align beams with the passive tag and thus increase the distinguishability of the tag's reflection signal.

\section{System Design}\label{sec:4}

In this section, we first show how to design a passive chipless  Wi-Fi tag for high accuracy soil moisture sensing. 
Next, we show how to improve the tag's work range by suppressing the LoS signal and multipath signal interference. 
Finally, we present a soil moisture estimation method.

\subsection{Passive Wi-Fi Tag Design}\label{sec:4.1}
In this section, we detail the design of our passive Wi-Fi tag for soil moisture sensing. The key is finding optimized parameters for a resonator whose frequency response is sensitive to even small soil moisture changes, thus resulting in a high sensing accuracy. We first show the resonator circuit and analysis the impact of soil moisture, then we detail how to optimize a resonator to improve the sensing accuracy.

\subsubsection{Resonator Circuit Design}

\begin{figure}[tb]
\setlength{\abovecaptionskip}{1mm}
\setlength{\belowcaptionskip}{-3mm}
\centering
\subfigure[3D structure of the resonator tag.]{
\label{fig:dgs3d}
\includegraphics[width=0.5\linewidth]{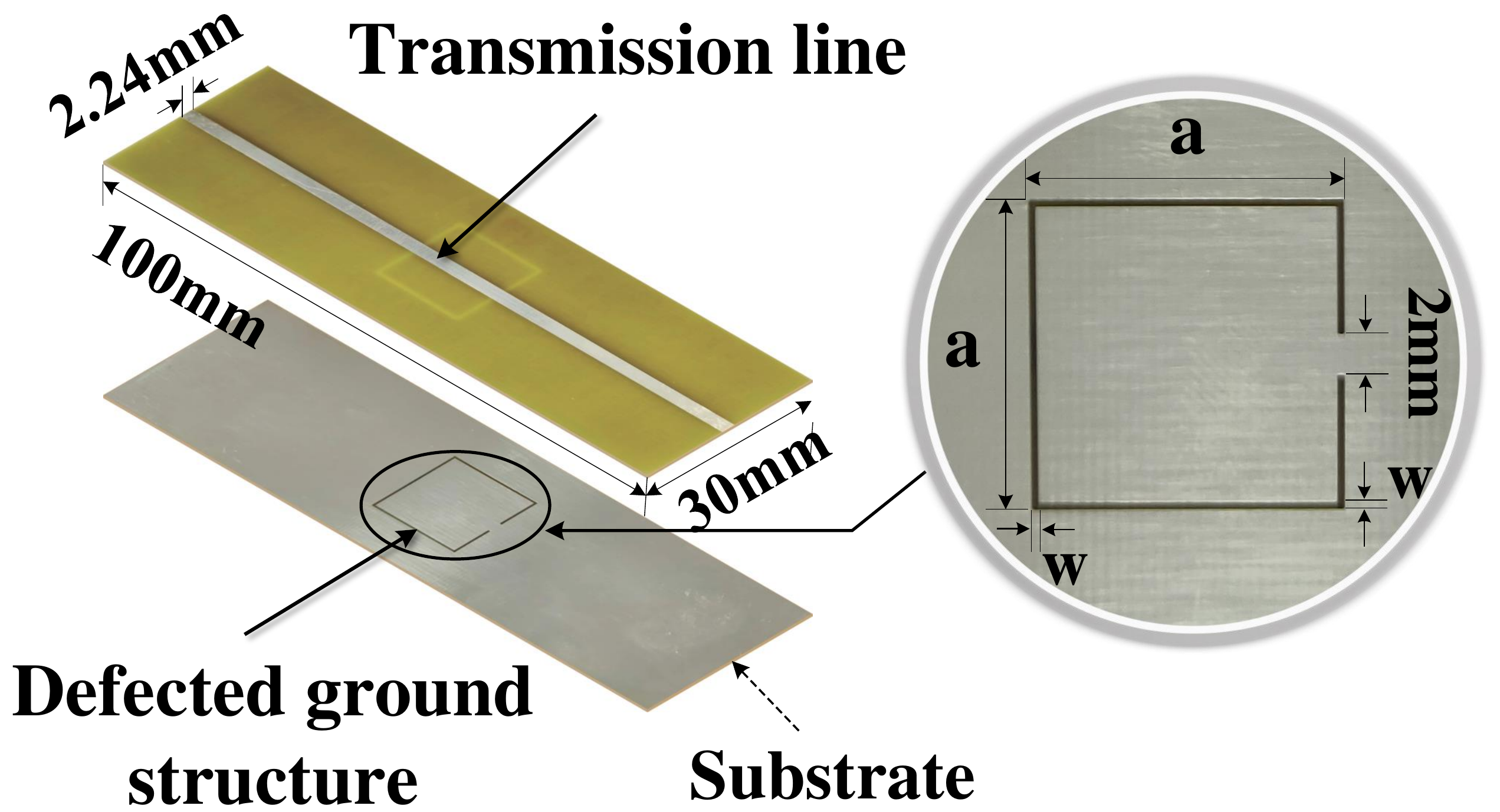}}
\hspace{1mm}
\subfigure[The equivalent circuit.]{
\label{fig:dgscircuit}
\includegraphics[width=0.44\linewidth]{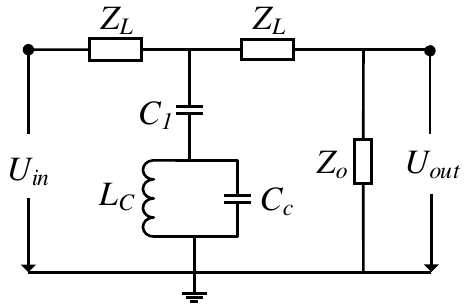}}
\caption{The geometric model and circuit model of the resonator tag.}
\end{figure}

The key component of our chipless Wi-Fi tag is a resonator, whose frequency response is a function of the inductance and capacitance of the circuit. When we attach the resonator to soil, the soil become a part of the circuit. 
Different soil moisture levels will have different soil permittivity and conductivity values, resulting in changes in the inductance and capacitance of the resonator circuit.  Thus, we aim to design a resonator circuit to produce different frequency responses when there are changes in the soil moisture.

To realize this design, we employ the microstrip circuit whose capacitance, resistance and inductance are distributed on the metal transmission line and the metal ground plane. To create a dynamic capacitor like a sliding rheostat, we use a special structure -- \textit{defected ground structure~\cite{DGS}}\footnote{DGS is a common structure used to form bandstop filter.}, which is an etched metal slot on the metal ground plane. The slot essentially forms a parallel resonant circuit. Fig.~\ref{fig:dgs3d} shows the 3-D structure of the microstrip resonant circuit. The microstrip transmission line, the metal ground plane and the four sides of substrate together form a resonant cavity. The shape and size of the defected ground structure determines the inherent resonant frequency of the circuit. When the soil contacts the defected ground structure, it will change the electromagnetic parameters (i.e., permittivity and conductivity) of the resonator, thereby changing its resonance frequency. When the signal passes through the resonator, the signal with different frequency component will go through different attenuation. Finally, the frequency spectrum of the output signal forms a feature to identify the soil moisture. In this way, we can transform different soil moisture to different frequency response features of the signal.

The microstrip resonance circuit can be equivalently modeled as a LC parallel resonance circuit shown in Fig.~\ref{fig:dgscircuit}. $Z_L$ is the equivalent resistor of the transmission line. $C_1$ represents the capacitor between the transmission line and the metal ground plane. $L_C$ and $C_C$ represent the equivalent inductor and capacitor of the defected ground structure. Mathematically, the frequency response function is the ratio between the output voltage $U_{out}$ and input voltage $V_{in}$:
\begin{small}
\begin{equation}\label{eq:frf}
\begin{array}{ccc}
FR(Z_{R}(\omega)) & = & 20 \log (\frac{U_{out}}{U_{in}}) \\
                  & = & 20 \log (\frac{Z_{o} \cdot Z_{R}(\omega)}{Z_{L}^{2}+2 \cdot Z_{L} \cdot Z_{R}(\omega)+Z_{o} \cdot Z_{R}(\omega)+Z_{o} \cdot Z_{L}}) \\
\end{array}
\end{equation}
\end{small}
where, $Z_{L}$ and $Z_{R}(\omega)$ are the impedance of the transmission line and defected ground structure, respectively, and $\omega$ is the angular frequency of the input signal. The $Z_{R}(\omega)$ can be expressed as:
\begin{small}
\begin{equation}\label{eq:imp}
Z_{R}(\omega) = \frac{1}{j \omega C_{c}+\frac{1}{j \omega L_{c}}}+\frac{1}{j \omega C_{1}}
\end{equation}
\end{small}
Eq.~\eqref{eq:imp} shows that the slot capacitor $C_C$ could change the impedance and thus change the frequency response formulated in Eq.~\eqref{eq:frf}. 

Based on Eq.~\eqref{eq:imp}, when the imaginary part of impedance reaches to zero, the resonance cavity reaches its resonance state. Then, the center frequency of the bandstop $f_c$ is:
\begin{small}
\begin{equation}\label{eq:fc}
f_{c} = \frac{1}{2\pi \sqrt{L_c(C_1+C_C)}}
\end{equation}
\end{small}

\subsubsection{Tuning Soil Resonator Parameters for working in Wi-Fi Band}
We show how to tune resonator parameters, so that the design resonator (or tag) can work in Wi-Fi band. For DGS structure resonator, the width and length of the defected structure determine the equivalent capacitance and inductor of the resonator, thereby determining its resonant frequency.
To design a proper resonator for soil moisture sensing, we should tune the geometric size of two components - the width of transmission line and the length and width of DGS structure. In our design, the thickness of substrate is given as $1.2$ $mm$.

Next, we elaborate our design process for a given soil moisture range and sensing accuracy. As illustrated in Fig.~\ref{fig:curve}, even if we can sense soil moisture from 0\%-100\%, it would loose the sensing accuracy in some soil moisture range (e.g., 10\%-100\%). Hence, to improve the sensing accuracy (e.g., 1\% accuracy) for the target soil moisture range, we need to adjust the sensitivity to soil moisture through optimizing the capacitance and inductance of the resonator. Given the soil moisture range $[\theta_1,\theta_2,...,\theta_n]$ corresponding to permittivity $[\epsilon_1,\epsilon_2,...,\epsilon_n]$, the optimization process needs two steps. 

First, we need to find the optimal parameter that make the resonant frequency of $\theta_1$ appear in the around of 2.5 GHz- 2.6 GHz. Specifically, we build a DGS geometry model in HFSS software and set the geometric parameters to make the resonant frequency larger than 2.48 GHz - the upper limit frequency of Wi-Fi 2.4 GHz band. For setting the initial geometric parameters, we give an simulation empirical parameters range for $w$ and $a$ in the next subsection. Then, we add a medium layer to the ground of the tag to simulate the effect of soil on DGS resonator and set the permittivity of this medium layer as $\epsilon_1$. Next, we use the parameter optimization function in HFSS software to search the optimal geometric parameters that can make the resonant frequency appear in 2.5 GHz - 2.6 GHz.

Second, we need to determine the optimal sensitivity. Specifically, we select the optimal geometric parameters of $w$ and $a$ from the obtained candidate values at first step to meet the sensitivity requirement. To do so, we build a new DGS model in HFSS and add a medium with $\epsilon_2$ permittivity corresponding to the second soil moisture $\theta_2$. Then, we can get the frequency offset $[\Delta f_1,\Delta f_2,...,\Delta f_k]$ and select the optimal frequency offset, which corresponds to a set of $w$ and $a$ values. 

\begin{figure}[t]
\centering
\setlength{\abovecaptionskip}{1mm}   
\setlength{\belowcaptionskip}{-3mm} 
\subfigure[$5\%$ sensing resolution in $0\%-20\%$ soil moisture. The geometric size $a=10mm$ and $w=0.1mm$.]
{\includegraphics[width=0.46\linewidth]{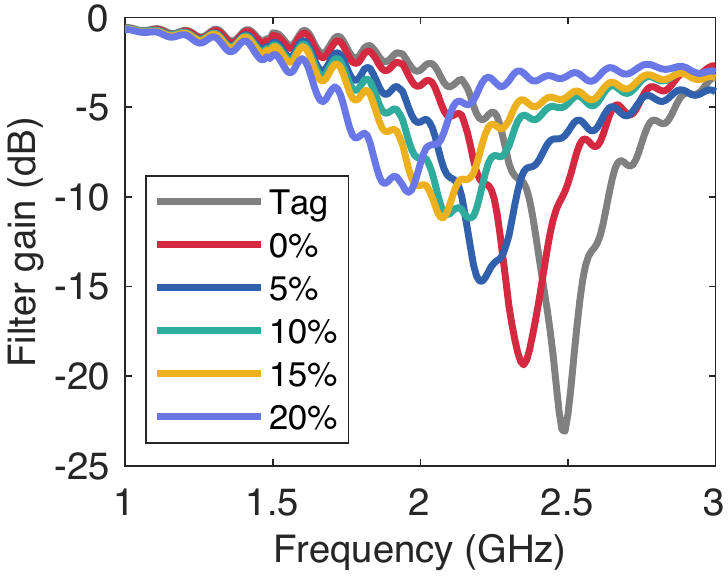}}
\hspace{3mm}
\subfigure[$2\%$ sensing resolution in $23\%-29\%$ soil moisture. The geometric size $a$=6 mm and $w$=1 mm.]
{\includegraphics[width=0.46\linewidth]{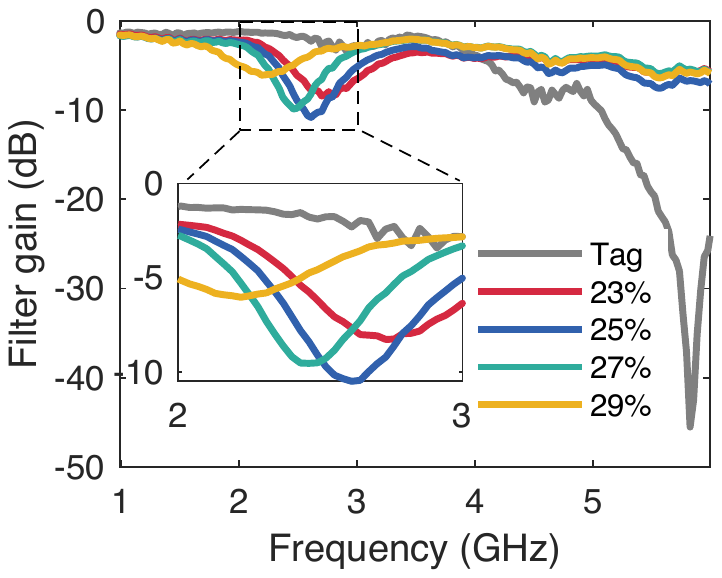}}
\caption{Frequency response features measured by VNA.}
\label{fig:s21}
\end{figure}

Following the two steps, we design two resonators with different sensing resolution. One resonator provides a relative high accuracy in small moisture range and the other provides a relative low accuracy in a wide moisture range. In order to validate that the two resonators can produce distinguishable frequency responses, we measure the frequency response using Vector Network Analyzer. The two ends of the tag are connected to VNA through RF cables. Fig.~\ref{fig:s21} shows the measured frequency response of two tags corresponding to soil moisture in $0\%-20\%$ and $23\%-29\%$. The resonator for sensing $0\%-20\%$ has 5\% resolution and the other for sensing 23\%-29\% has 2\% resolution. This result demonstrates the effectiveness of our design to customize the sensing resolution and soil moisture range.

\subsubsection{The impact of DGS size on resonant frequency}
\begin{figure}[!t]
\setlength{\abovecaptionskip}{1mm}   
\setlength{\belowcaptionskip}{-3mm}  
\centering
\subfigure[The width $w$.]{
\label{fig:w}
\includegraphics[width=0.47\linewidth]{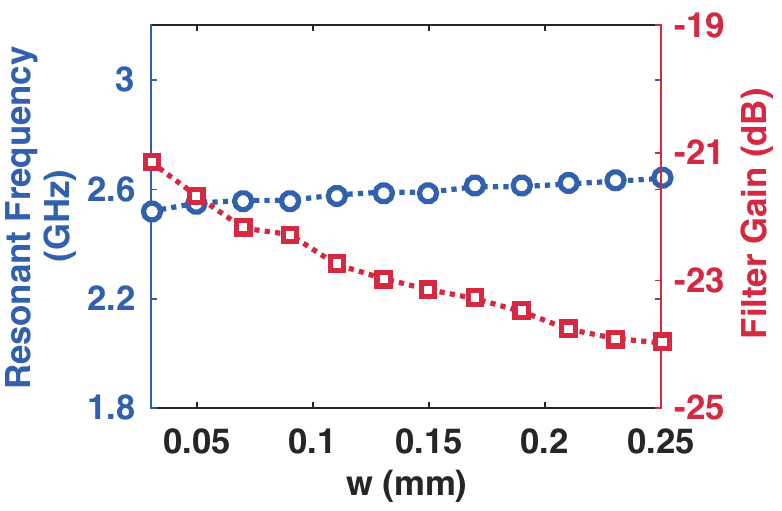}}
\hspace{1mm}
\subfigure[The length $a$.]{
\label{fig:a}
\includegraphics[width=0.47\linewidth]{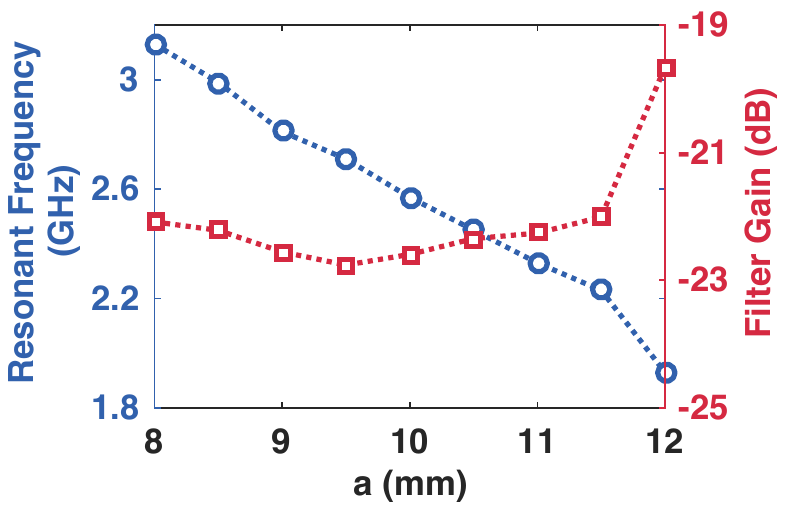}}
\caption{The impact of the defected structure size on the resonant frequency and filter gain.}
\label{fig:wa}
\end{figure}

We analyze the impact of width $w$ and length $a$ on the resonant frequency and filter gain through HFSS simulation. Fig.~\ref{fig:wa} shows the simulated results about how the width and length of the resonance structure affect the resonant frequency and filter gain. 

\noindent \textit{(i) The slot width $w$.} Fig.~\ref{fig:w} indicates that a wider slot leads to higher resonance frequencies, because it increases the distance between two plates of the capacitor which essentially decreased the capacitance value $C_C$. Therefore, the resonant frequency gets higher.

\noindent \textit{(ii) The slot length $a$.} Fig.~\ref{fig:a} shows that as $a$ increases, the resonant frequency shifts towards lower frequencies. This is because a longer slot means larger area of two plates of the capacitor $C_C$ and also leads to a larger capacitance value. In addition, the length $s$ has a slight impact on the filter gain.

Based on the above results, one should set the width $w$ as 0.1~mm and the length $a$ as 10$\sim$11 mm, so that the resonator can work in Wi-Fi band.

\subsection{Passive Beam Alignment for Long  Range}\label{sec:beam}

Usually, the communication range of passive Wi-Fi devices is limited due to the strong interference signals from the transmitter’s direct path or multipath. To improve the passive Wi-Fi tag's working range, we propose a passive beam alignment method so that the interference signals can be suppressed and the signal-to-noise ratio (SNR) of the tag's backscatter signal can be boosted at the receiver.

\begin{figure}[tb]
\centering
\setlength{\abovecaptionskip}{1mm}
\setlength{\belowcaptionskip}{-3mm}
\includegraphics[width=0.7\linewidth]{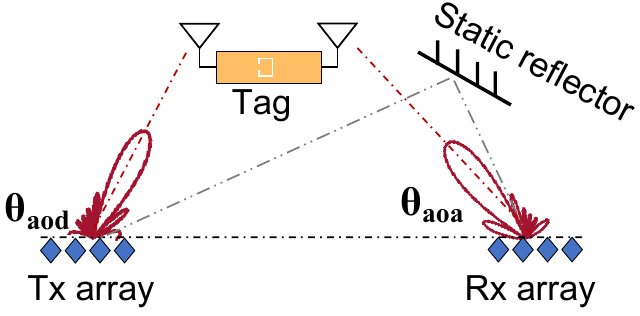}
\caption{The transceiver pair reduce the LoS and multipath interference by aligning their beams with the tag.}
\label{fig:beamalign}
\end{figure}

Our key idea is that:  if the transmitter and receiver can point their beams to the passive tag rather than the background reflectors, as shown in Fig.~\ref{fig:beamalign}, then we can strength the tag's incident signal and backscatter signal by performing beam-forming at both the transmitter and receiver side. However, the challenge is that the tag has no battery and is unable to perform channel estimation. Thus, it is impossible to let the tag to feedback its spatial angles to the transmitter and receiver.

\begin{figure}[tb]
\centering
\setlength{\abovecaptionskip}{1pt}
\setlength{\belowcaptionskip}{-3mm}
\centering
\subfigure[AoD profiles measured by receiver.]{
\label{fig:aod}
\includegraphics[width=0.45\linewidth]{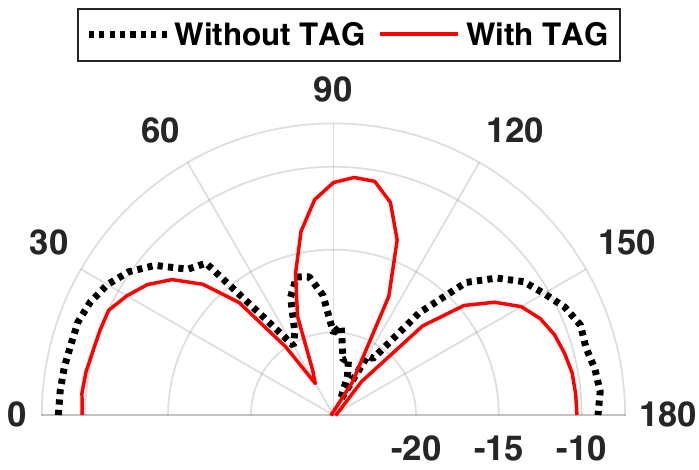}}
\hspace{3mm}
\subfigure[AoA profiles measured by MUSIC algorithm.]{
\label{fig:aoa}
\includegraphics[width=0.45\linewidth]{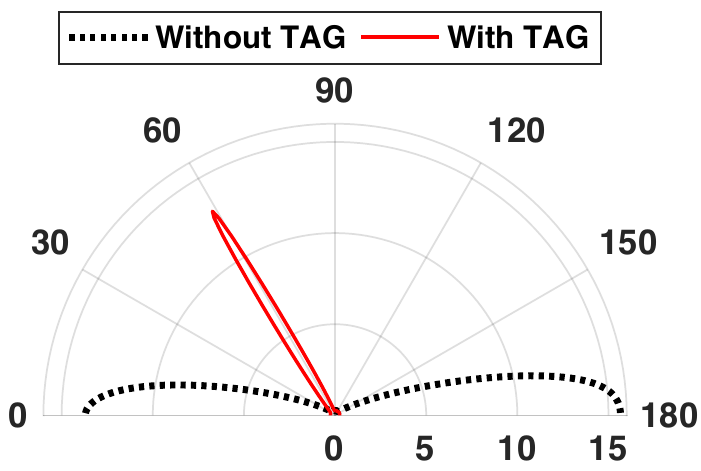}}
\caption{Estimating the direction of tag relative to the transmitter and receiver by comparing two spatial spectrum profiles with tag and without tag.}
\label{fig:locate}
\end{figure}

To solve this challenge, we leverage the receiver to estimate the relative angle of the tag to Tx $\theta_{aod}$ and Rx $\theta_{aoa}$, as shown in Fig.~\ref{fig:beamalign}. For the estimation of $\theta_{aod}$, we leverage the ability of beam scanning of the transmitter. Specifically, the transmitter steers its beam at a uniform speed while the receiver measures the power of each received packet and finally synthesize the spatial power profile corresponding to Tx's beam direction in  $0^{\circ}-180^{\circ}$. To estimate $\theta_{aod}$, we leverage an observation: when the transmitter's beam points at Rx and the tag, the power profile could have the largest and a second largest peak, respectively. By finding the angle when the second largest peak appears, the $\theta_{aod}$ can be obtained. Fig.~\ref{fig:aod} shows two spatial power profiles for without tag and with tag. From the red profile, we can determine there is a tag at about $102^{\circ}$ direction because this angle is the second peak in this profile.

To estimate $\theta_{aoa}$, we use the antenna array at the receiver and the MUSIC algorithm~\cite{joshi2014direction,laxmikanth2015enhancing}. Specially, the peak in the MUSIC estimated AoA spectrum shows the angle values of $\theta_{aoa}$. After obtaining $\theta_{aod}$, the receiver estimate the AoA by using the signal phase captured at the timing when the transmitting signal point to the tag (i.e., the timing appearing the second power peak in AoD profile). Fig.~\ref{fig:aoa} shows two spatial power profile by MUSIC. By comparing the two profile, we can find the tag at $60^{\circ}$ direction. 

\begin{figure}[t]
\setlength{\abovecaptionskip}{1mm}
\setlength{\belowcaptionskip}{-3mm}
\centering
\subfigure[The fabricated microstrip patch array antenna.]{
\label{fig:array}
\includegraphics[width=0.4\linewidth]{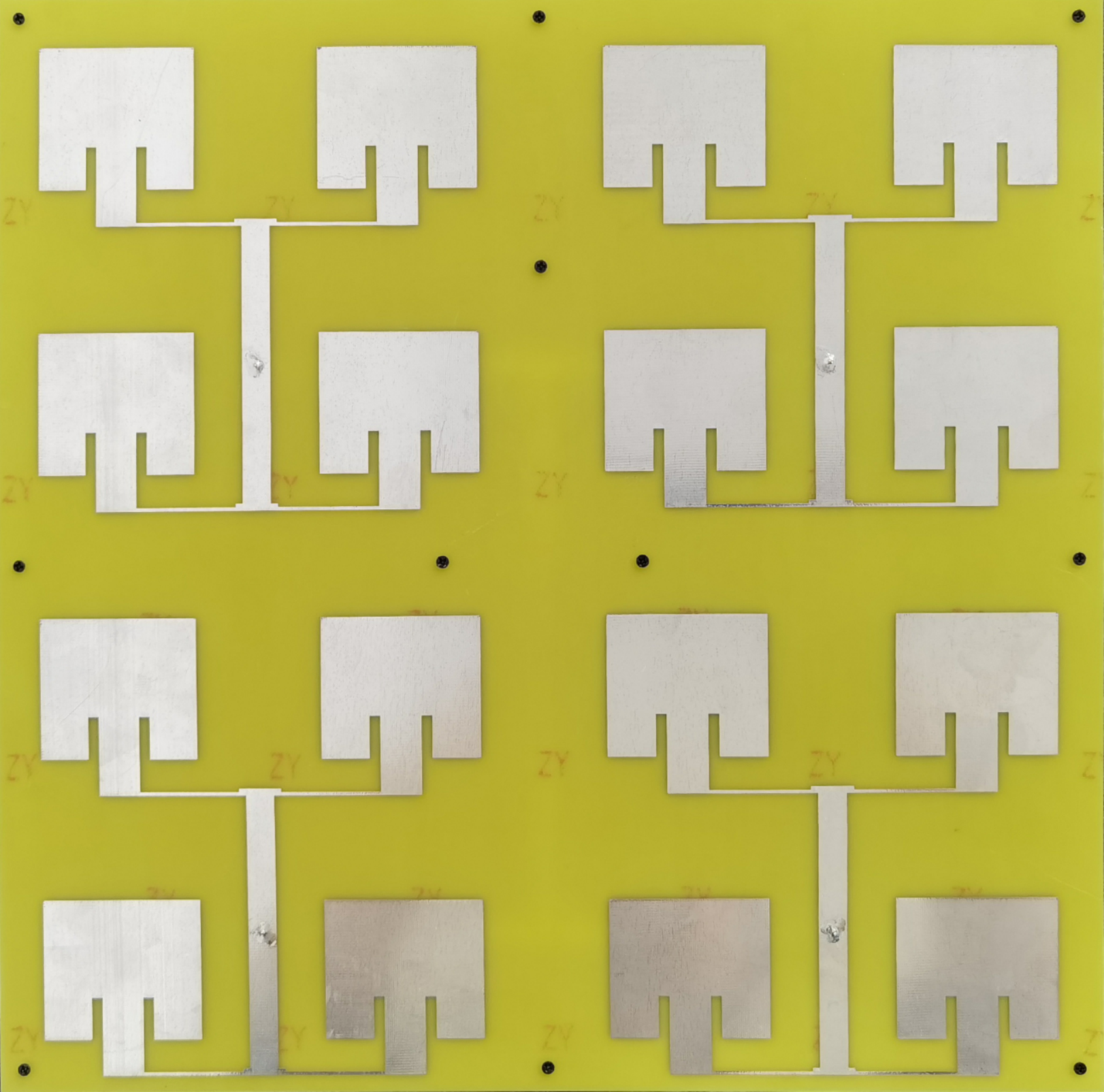}}
\hspace{5mm}
\subfigure[The radiation pattern of 16-elements patch array.]{
\label{fig:radpat_array}
\includegraphics[width=0.4\linewidth]{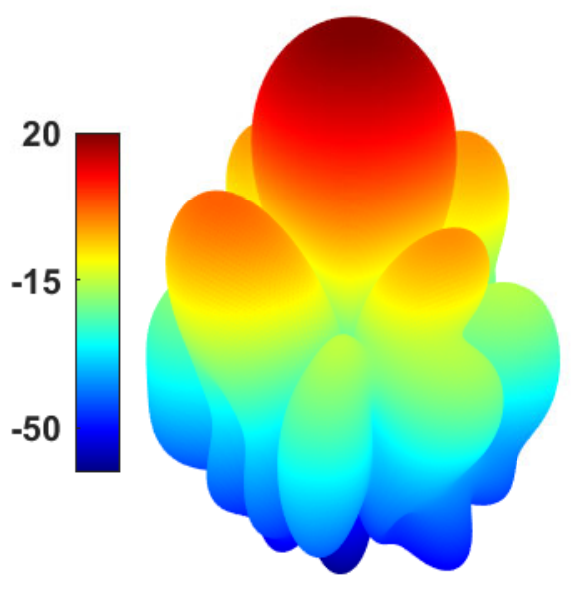}}
\caption{The fabricated high gain microstrip antenna for the passive tag.}
\end{figure}

Finally, we design high gain antennas for the passive tag to further strengthen the tag's backscatter signal. A simple design is using multiple patch antenna units as shown in Fig.~\ref{fig:radpat_array}. The more patch antenna units are, the higher the antenna gain is, but it also results in a larger antenna size. To trade-off between the antenna gain and the size, we use four patch antenna array in this paper. The designed 4$\times$4 patch array can achieve 16.8~dBi radiation gain, as shown in Fig.~\ref{fig:radpat_array}.

\subsection{Soil Moisture Estimation} \label{sec:Feature}
\subsubsection{CSI feature extraction}
The distortion of wireless channel caused by the tag manifests in the channel-state-information (CSI). To capture the channel frequency response feature, the transmitter sends packet over 13 channels and the Wi-Fi receiver estimates the channel state information for each transmission packet. By splicing CSIs of 13 channels, we can obtain the frequency response feature.

To do so, existing approaches leverage the fact that Wi-Fi has three non-overlapping channels \cite{Bejarano2013IEEE}. By stitching CSI of the three non-overlapping channels, one can obtain an approximate frequency response curve. However, this approximate frequency response curve could miss the information of some frequency points, making two profiles indistinguishable. This is because two frequency response curves may overlap in a certain channel due to noise or multipath interference. Fig.~\ref{fig:csifeature} shows nine CSI profiles of nine soil moisture levels in $0\%-20\%$. We can see CSI profiles of two adjacent soil moisture levels may overlap at some frequency points. If we only see the frequency response of these frequency points, the two curves are indistinguishable. When we look the two curves in the whole Wi-Fi band, the two profiles can be distinguishable. Therefore, we should splice the CSIs covering 2.4GHz-2.48GHz. Specifically, we select a little consecutive subcarriers from each channel and arrange all the selected subcarriers in frequency order. Finally, all of the selected subcarriers can form a frequency spectrum feature.

\subsubsection{Removing the impact of distance on the CSI amplitude variations}

In reality, the distance changes between Wi-Fi transceivers and the tag also could change the power amplitude, resulting an error if we use the absolute  amplitude values. To address this problem, we obtain the relative amplitude attenuation ratio rather than the absolute amplitude of the signal. Specifically, we fist measure a reference CSI amplitude $A_{ref}$ before attaching the tag to soil. Then, when the tag is attached to the soil, we get another online measured CSI amplitude $A$. Next, we calculate the ratio between the pre-measured reference amplitude and the online measured amplitude. We define this ratio as filter gain:
\begin{equation}
\text{Filter gain (dB)}  =  20\log\frac{A}{A_{ref}}.
\end{equation}

\begin{figure}[t]
\setlength{\abovecaptionskip}{1mm}
\setlength{\belowcaptionskip}{-4mm}
\centering
\subfigure[CSI features across WiFi band.]{
\label{fig:csifeature}
\includegraphics[width=0.5\linewidth]{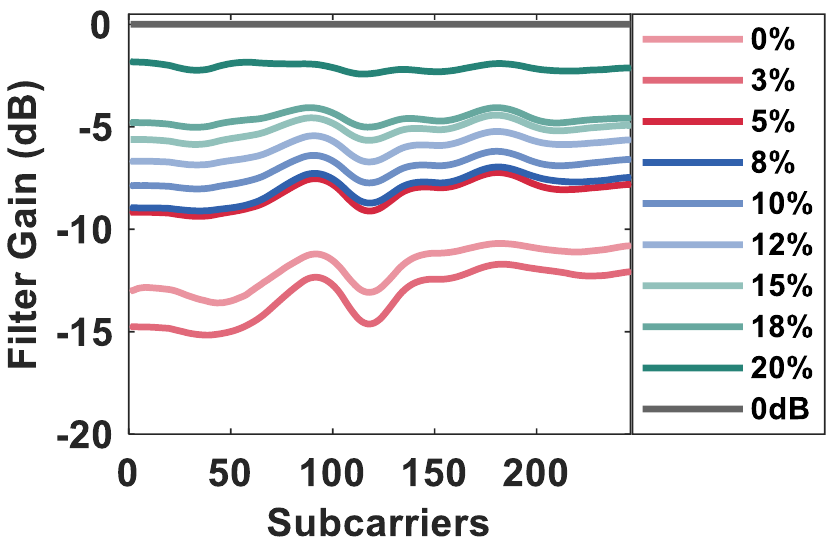}}
\hspace{3mm}
\subfigure[CSI power feature of soil moisture.]{
\label{fig:csipower}
\includegraphics[width=0.4\linewidth]{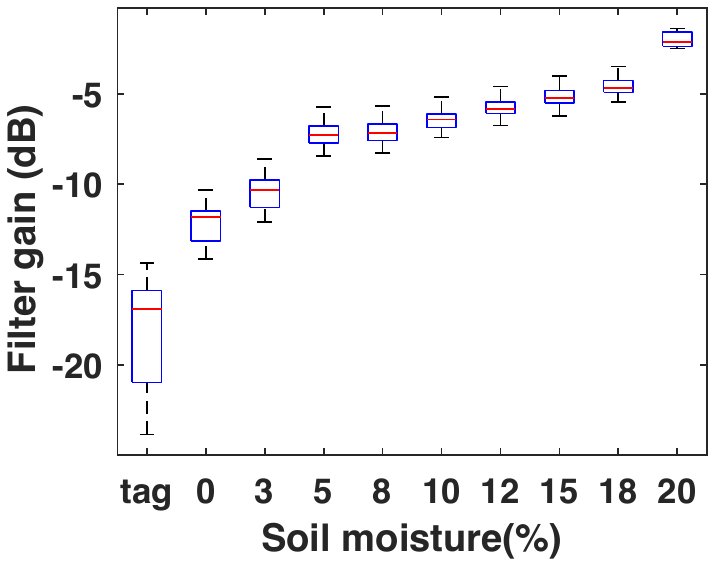}}
\caption{The channel frequency response features and CSI power feature of different soil moisture levels.}
\end{figure}

Fig.~\ref{fig:csipower} plots the mean filter gain of each CSI feature profile in Fig.~\ref{fig:csifeature}. We can see that the filter gain increases with the soil moisture increasing. This result demonstrates the effectiveness of CSI amplitude ratio to sense humidity.

\subsubsection{Soil-moisture Estimation}
After obtaining the CSI profiles of different soil moisture, SoilTAG needs to match the profile of the unknown soil moisture with the profiles of known soil moisture in order to infer soil moisture level.
A simple approach is dynamic time wrapping (DTW), which can compute the similarity of two curves. To identify the soil moisture, DTW algorithm measures the similarity of the unknown CSI feature and all of the known CSI features of different soil moisture levels.
However, the CSI feature of the same soil moisture level may show a difference in amplitude and the curve shape between two measurements, due to multipath changes in the environment, or slight changes in the degree of soil compaction.
The instability of soil moisture feature will cause the soil moisture value estimated by DTW algorithm to deviate greatly from the real soil moisture value.
To address this problem, we employ random forest algorithm because of two advantages anti-instability and outstanding generalization performance. To achieve good generalization performance, the training set should be diverse. Thus, we collect $33200$ traces of CSI feature data of different soil moisture in five environments and use half of traces as the training set and the other traces as the test set.

\section{Evaluation}
\subsection{Experimental Method}
\begin{figure}[!t]
\setlength{\abovecaptionskip}{1mm}
\setlength{\belowcaptionskip}{-3mm}
\centering
\includegraphics[width=0.98\linewidth]{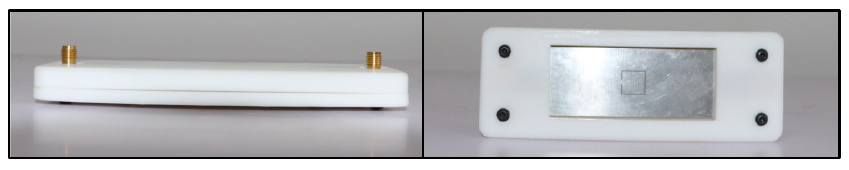}
\caption{The fabricated chipless tag.}\label{fig:tag}
\end{figure}

\begin{figure}[!t]
\setlength{\abovecaptionskip}{1mm}
\setlength{\belowcaptionskip}{-5mm}
\centering
\subfigure[The soil samples prepared by dry weighing method.]{\label{fig:loam_soil}{
\includegraphics[width=0.63\linewidth]{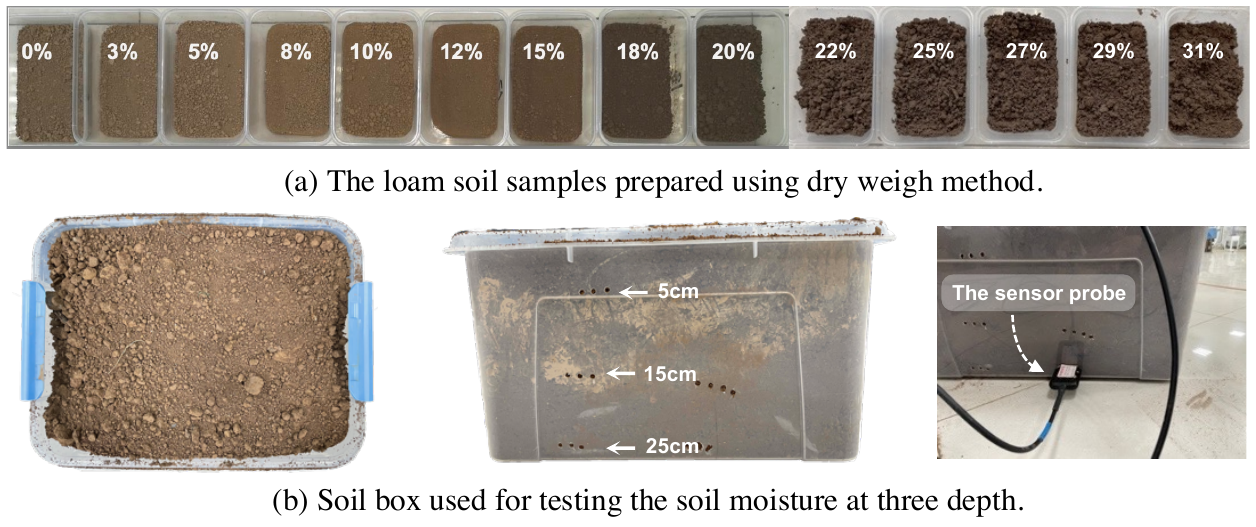}}}
\hspace{1mm}
\subfigure[Delta-T WET-2 Kit.]{\label{fig:delta-t}{
\includegraphics[width=0.3\linewidth]{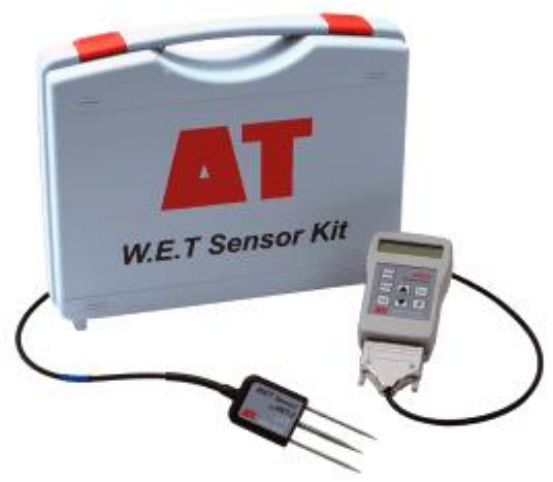}}}
\caption{Two methods are used as the ground truth.}
\end{figure}

\textbf{Implementation.} Our \systemname system consists of three components: the chipless tag, Wi-Fi transceiver, and the backend soil moisture estimation software.

\textit{Chipless tag.} The chipless tag is comprised of a DGS resonator and two antennas connectors as. 
To produce the resonator, we first design the DGS resonator of chipless tag by the design method proposed in Sec.~\ref{sec:4.1}. A laminate board consists of two copper layers and the dielectric substrate in between. We employ standard PCB milling technology to fabricate the 2D layout of the conductive layer on one side of the laminate, whereas the back side is directly used as the ground layer. The substrate is low-cost FR4 material with a thickness of 1.2~mm. The width of transmission line is 2.24~mm. The two antennas are connected with the DGS resonator through two RF cables. To sense soil moisture, the DGS resonator can be buried into deep soil or attached on the soil surface and the two antennas are placed on the ground for receiving and reflecting the signal.

\textit{Transceiver.} We employ two types of devices as the transceiver. To implement the beam-forming mechanism, we use the WARP software radio which works at 2.4~GHz with 80~MHz spectrum. The transmitter is configured with 8-antennas linear array to perform beam scanning and the receiver is configured with 4-antennas linear array to realize the tag's angle estimation introduced in Sec.~\ref{sec:beam}. The transmit power levels is set lower than 30~dBm. We implement the beam-scanning at transmitter based on this work~\cite{sur2016practical}.

\textit{Backend implementation.} Our data process and soil moisture estimation algorithm are implemented in MATLAB and Python code respectively. We use a laptop with a 2.3 GHz CPU (Intel i7-10875H) and 16~GB memory to run our software. The laptop is connected to WARP radios through a switch and Ethernet cables, and they communicate using WAR-PLab mode~\cite{url8}.

\begin{figure}[t]
    \centering
    \begin{minipage}[!t]{1\linewidth}
    \setlength{\abovecaptionskip}{1mm}
    \setlength{\belowcaptionskip}{-5mm}
    \centering
    \includegraphics[width=0.4\linewidth]{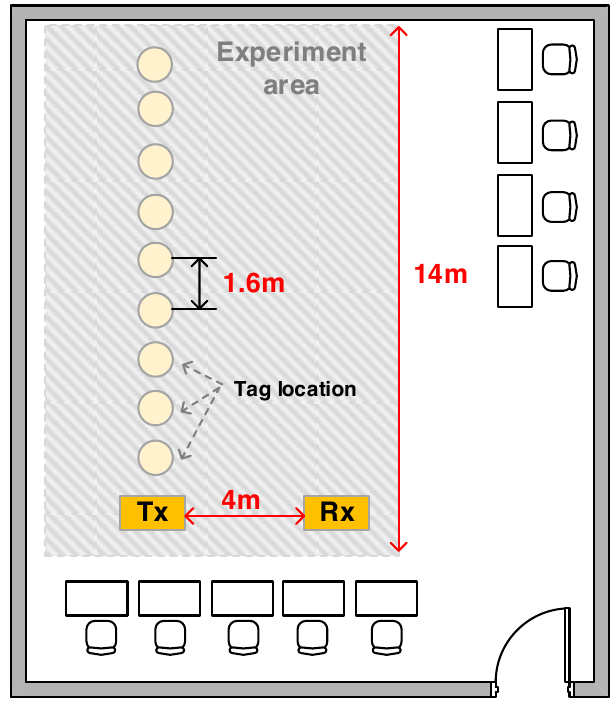}
    \caption{The experiment layout.}\label{fig:layout}
    \end{minipage}
\end{figure}

\textbf{Experimental setup and deployment.} The default setup is attaching the tag to different soil samples. When evaluating the sensing depth underground in Sec.~\ref{sec:evaluation:depth}, we bury the tag into the soil and put the soil box in the experimental area. The experiment setup is shown in Fig.~\ref{fig:layout}.
The transmitter (Tx) and receiver (Rx) are placed with a separation of 4 m.

\textbf{Evaluation metrics.} The soil samples used in experiments are the loam soil, which is very porous and retains moisture well, the optimal soil type, shown in Fig.~\ref{fig:loam_soil}.
The sensing accuracy is defined as the absolute error between the estimated soil moisture and the truth soil moisture. 
Except for the deep soil moisture evaluation, the ground-truth is measured using the dry weighing method. The soil samples are shown in Fig.~\ref{fig:loam_soil}a.
To evaluate the sensing depth, the tag is buried into the deep soil and thus the ground truth at the tag's location is measured by a dedicated soil moisture measurement device, i.e., Delta-T WET-2 Kit device~\cite{urle}. To remove the measurement error, we set the soil type of the WET-2 meter to the loam soil which is pre-defined in the WET-2 meter. The unit of soil moisture sensor's reading value is the volumetric water content (VWS\%). We convert the sensor's reading to the gravimetric water content (GWS\%) by dividing the soil density.

\textbf{Experiment road map.} We first evaluate the overall sensing accuracy of \systemname and compare our system with existing soil moisture sensors and systems. Next, we evaluate the impact of several factors on the sensing accuracy in Sec.~\ref{sec:evaluation:factors} including $(i)$ the effect of the patch array antenna, $(ii)$ the effect of the beam forming, 
$(iii)$ the depth of tag buried into the soil, $(iv)$ the impact of tx-tag distance, $(v)$ the impact of human motion around the system, $(vi)$ the impact of different environments). 
Finally, we conduct a case study to evaluate the performance of \systemname through a continuous  soil moisture monitoring.

\subsection{Overall Accuracy of Soil Moisture Sensing}\label{sec:evaluation:accsoil}

We evaluate the overall accuracy of \systemname for soil moisture sensing.
The test data of different soil moisture is collected  in an empty room.
In this experiment, we configure the transceiver with the directional antenna, which has $60^{\circ}$ beam width and 16.8~dBi gain. The transmit antenna and the receive antenna are placed with a separation of 4~m. The chipless tag is placed 0.5~m away from the transmit and receive antennas.
We evaluate \systemname when the soil moisture level changes from 0\% to 90\%.
\footnote{We do not test moisture levels that are higher than 90\%, because a lot of water overflows on the soil surface in this case. Thus the data is less useful in reality.} 
The ground truth is measured by a dedicated soil moisture measurement device, i.e., Delta-T WET-2 moisture meter.

\begin{figure}[!t]
\setlength{\abovecaptionskip}{0mm}
\setlength{\belowcaptionskip}{-3mm}
\centering
\begin{minipage}[!t]{1\linewidth}
\centering
\subfigure[The soil moisture estimation results.]{
\label{fig:predict}{
\includegraphics[width=0.46\linewidth]{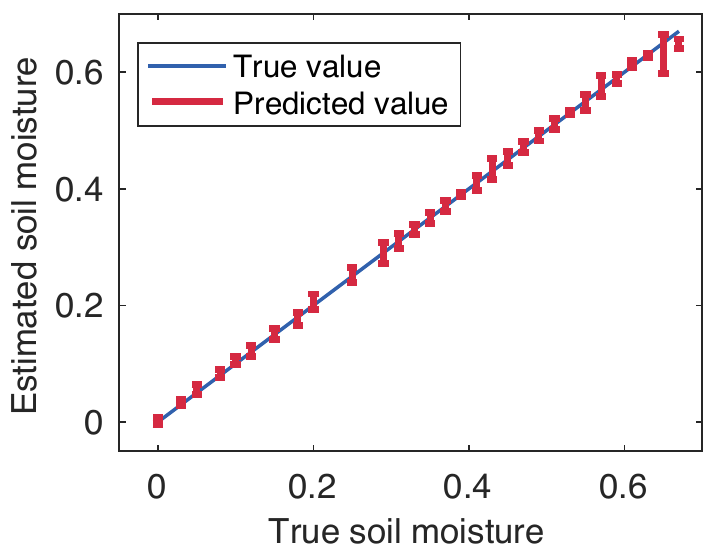}}}
\hspace{1mm}
\subfigure[CDF of  moisture estimation errors.]{
\label{fig:acc_cdf}
\includegraphics[width=0.46\linewidth]{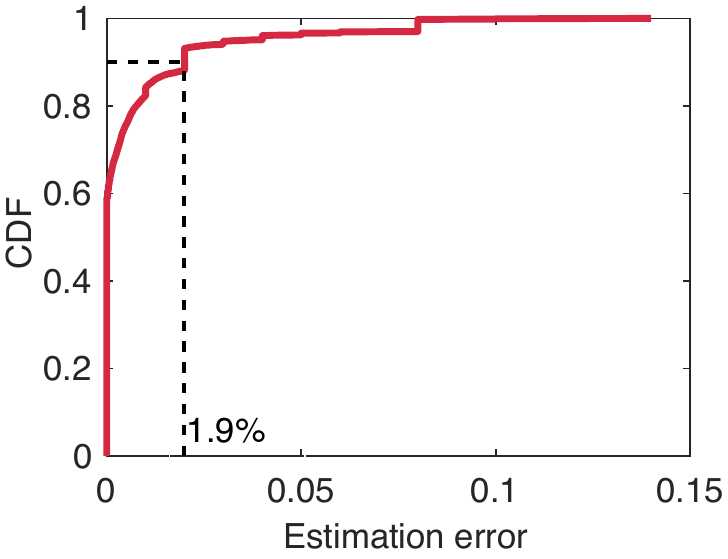}}
\caption{The overall soil moisture estimation accuracy.}
\end{minipage}
\end{figure}

For each soil moisture level, we collect 100 data packets and extract the relative frequency response features. Then, we input these features into a well-trained regression model to estimate the soil moisture. Fig.~\ref{fig:predict} shows the estimated soil moisture value and the ground truth. As we can see, the maximum variance of estimation error is only $\pm$ 2\%. We compute the cumulative distribution function (CDF) of estimation errors, which is plotted with the red curve in Fig.~\ref{fig:acc_cdf}. We find that the $90th$ estimation error is 2\%.

In addition, we comprehensively compare \systemname system with existing methods in the sensing accuracy, working range and the system price, as shown in Table. 1. It shows that our \systemname not only achieves the high accuracy and a long working range but also the cheap cost. The overall cost of \systemname includes the cost of two commodity Wi-Fi cards, the tag and two patch array antennas. The price of tag and patch array antenna are the unit price of producing ten thousand units.

\begin{table*}[!t]\footnotesize
\centering
\setlength{\tabcolsep}{5pt}
\renewcommand{\arraystretch}{0.8}{\multirowsetup}{\centering}
\caption{Comparison with existing moisture sensors/systems.}\label{tb:sensor}

\begin{tabular}{l|l|l|l|l|l}

\toprule

\textbf{Soil Moisture Sensor}  & \textbf{Accuracy} & \textbf{Working Range} & \textbf{Measurable Scale} & \textbf{Max Depth}  & \textbf{Price (USD)} \\

\midrule

\rowcolor{Gray} IC-T-350~\cite{urlc} & $\pm 1.5\%$ & Handheld & 0\%-100\% & 15~cm & 2399.00 \\

                IC-MPKit-406B~\cite{urlb} & $\pm 2\%$ & Handheld & 0\%-100\% & 15~cm & 1999.00  \\

\rowcolor{Gray} Delta-T Wet-2~~\cite{urle}& $\pm 2\%$ & Handheld & 0\%-100\% & 15~cm & 1904.00  \\

                PMS-714~\cite{urld} & $\pm 5\%$ & Handheld & 0\%-100\% & 15~cm & 185.44 \\

\rowcolor{Gray} ICZD06~\cite{urla} & 12.5\% & Handheld & 0\%-50\% & 15~cm & 99.00  \\

                Strobe~\cite{ding2019towards} & $\pm 10\%$ & 1.5~m & 0\%-100\% & 15~cm & 90.00  \\

\rowcolor{Gray} GreenTag~\cite{wang2020Soil} & $\pm 5\%$ & 2~m & 0\%-100\% & 0~cm & 1500.00 \\

                In-ground~\cite{josephson2021low} & $\pm 3.4\%$ & 3~m & 0\%-35\% & 60~cm & 56  \\
                
\rowcolor{Gray} {\systemname} (Our system) & $\pm 2\%$  & 6~m & 0\%-90\% & 1~m+ & 50.39  \\
\rowcolor{Gray}            & $\pm 3.64\%$ & 10~m   & 0\%-90\% & 1~m+ & 50.39 \\
\rowcolor{Gray}            & $\pm 8\%$    & 13.9~m & 0\%-90\% & 1~m+ & 50.39 \\

\bottomrule
\end{tabular}
\end{table*}

\subsection{Impact of System Parameters and Environment Noise}\label{sec:evaluation:factors}
We discuss the impact of different system parameters (e.g., the tag's deployment depth and range) and the environmental noise (e.g., moving object around the setup) on the sensing accuracy.

\subsubsection{The impact of tag's antenna on working range}
\label{sec:evaluation:antenna}
To boost the tag's reflection signal, we designed a patch array antenna for the tag. In this experiment, we aim to evaluate the effect of patch array antenna on improving the working range. Specifically, we compare the patch array antenna against a commodity omnidirectional dipole antenna with 5 dBi gain. This experiment is conducted in an empty corridor and the soil sample with different moisture levels are placed in this experimental area. The main beams of transmitter and receiver point to the tag's two antennas respectively. The Tx-Tag distance changes at a step size of 1m.

\begin{figure}[!t]
\setlength{\abovecaptionskip}{1mm}
\setlength{\belowcaptionskip}{-6mm}
\centering
\begin{minipage}[!t]{0.48\linewidth}
\setlength{\abovecaptionskip}{1mm}
\setlength{\belowcaptionskip}{-6mm}
\centering
\includegraphics[width=1\linewidth]{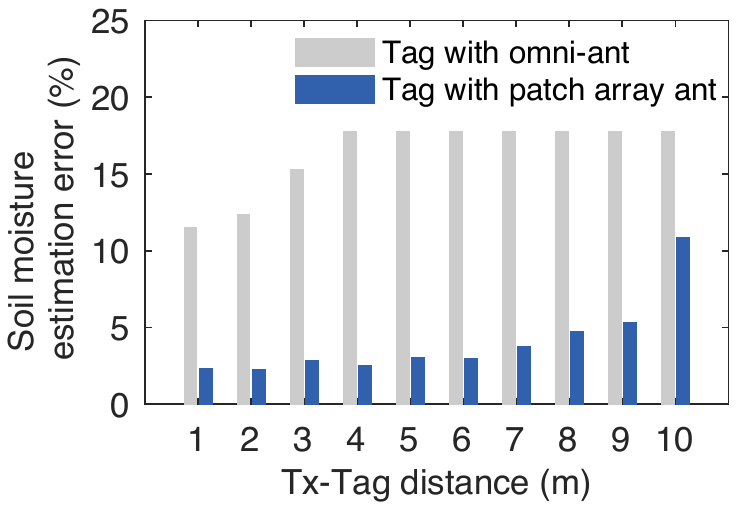}
\caption{The impact of the patch array antenna.}
\label{fig:tag_ant_gain}
\end{minipage}
\hspace{1mm}
\begin{minipage}[!t]{0.46\linewidth}
\setlength{\abovecaptionskip}{1mm}
\setlength{\belowcaptionskip}{-6mm}
\centering
\includegraphics[width=1\linewidth]{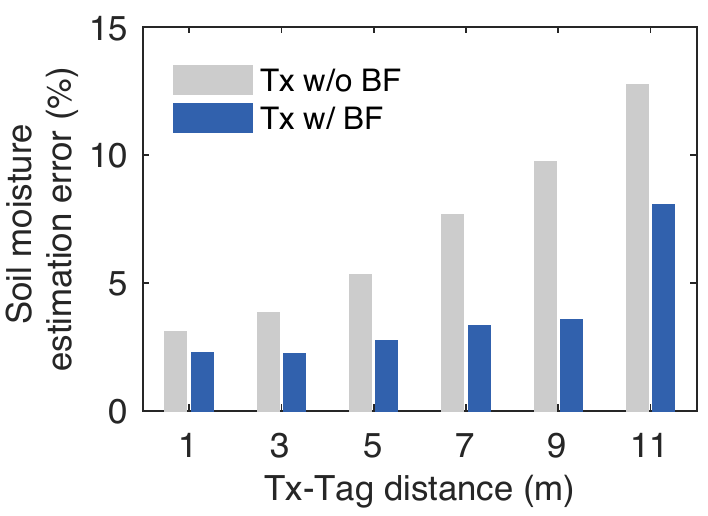}
\caption{The impact of beam forming.}
\label{fig:beam_align_gain}
\end{minipage}
\end{figure}

Fig.~\ref{fig:tag_ant_gain} reports how the sensing error varies with distance when the tag is equipped with the two types of antennas respectively.
For the omnidirectional antenna, we can see that even within 1~m Tx-Tag distance, the sensing error is more than 10\%. When the Tx-Tag distance is over 3~m, the sensing accuracy gets worse because the relative frequency response features are indistinguishable.
In contrast, when the tag configured with the patch array antenna, the working range can achieve 7~m that keeps the estimation error less than 5\%.
This is because the received signal and the reflection signal with omnidirectional antenna are much weaker than that of patch array antennas. 
When the tag's reflection signal gets weak and even reaches to the noise floor, the relative frequency response would greatly deviate from the true value, and hence resulting in the increase of sensing error.

\subsubsection{The effect of the beam-forming on improving the working range}
To evaluate the effect of the beam-forming mechanism on improving the working range, we compare the sensing error with and without the beam-forming.
This experiment is conducted in an empty corridor.
The tag is configured with the patch array antenna. The tag is placed in front of the transmitter's antenna array ($90^\circ$ direction) and the transmitter's main beam is fixed at $90^\circ$ direction. The Tx-Tag distance increases from 1~m to 11~m at a step of 2~m.

Fig.~\ref{fig:beam_align_gain} reports the sensing error under different Tx-Tag distance. We find that in the case of without beamforming, the working range that the sensing error keeps lower than 5\% is 4~m. 
In contrast, in the case of using beamforming, the working range can reach 9~m, has an improvement by 1.25$\times$. 
This result demonstrates the effectiveness of the Tx beamforming on improving the working range. In addition, we can see that the sensing error increasing with Tx-Tag distance increasing. 
The reason are twofold. One is that the received signal strength of the tag gets weaker when moving the tag away from Tx. The other is the presence of LoS signal leakage from Tx's side lobes to Rx, which is inevitable in beamforming technology. Even with side lobe's leakage, the working range is larger than existing RF-based soil moisture sensing system.

\subsubsection{Impact of the deployment depth}\label{sec:evaluation:depth}
The sensor tag may be buried into the soil in real-world applications. Thus, we evaluate the impact of different deployment depth levels on the sensing accuracy. 
We bury three tags at depth levels of 5~cm, 15~cm and 25~cm. The tag has a resin material coat to prevent it from shorting circuit by the soil/water, as shown in Fig.~\ref{fig:tag}. The distance between the two transceivers is 4~m. The Tx-to-Tag distance is 1~m. The system automatically collects data every half an hour. At the same time, we measure the ground truth of soil moisture with the Delta-T WET-2 moisture meter. The soil box used in this experiment is shown in Fig.~\ref{fig:loam_soil}b is placed in this experimental area.

Fig.~\ref{fig:depth_cdf} shows the CDF of soil moisture estimation errors in three depth levels. We can see that \systemname achieves the an accuracy of 1.4\% at a depth of 25~cm and 2.58\% at a depth of 15~cm. However, the accuracy decreases to 9.9\% at the depth of 5~cm. 
By observing the measurements, we discovered that our system's soil moisture estimate values are rather steady at three depths, whereas the sensor's readings at the depth of 5 cm are lower. As a result, the estimation error at a depth of 5 cm seems to be higher than 15 cm and 25 cm. The reason is that deep soil moisture is more homogeneous and steady. This means that both the tag and the sensor must be in full contact with the soil in order to acquire a more accurate soil moisture measurement in practice.

\subsubsection{Impact of the deployment range}\label{sec:evaluation:distance}
The tags may be deployed at various ranges away from transceivers, which may affect the sensing accuracy. Thus, we discuss the impact of a tag's working range on moisture sensing accuracy. 
The transmitter and receiver are placed with a separation of 4 meters.
We fix the Rx-Tag distance and change the Tx-Tag distance from 0.3~m to 13.9~m with a step size of 0.8~m by moving the tag's receiving antenna. 
At each distance, we collect 500 packets for each soil moisture levels and calculate estimation errors.

\begin{figure*}[!t] 
\setlength{\abovecaptionskip}{1mm}
\setlength{\belowcaptionskip}{-5mm}
\centering
\begin{minipage}[!t]{0.31\linewidth}
\setlength{\abovecaptionskip}{1mm}
\setlength{\belowcaptionskip}{-5mm}
\centering
\includegraphics[width=0.95\linewidth]{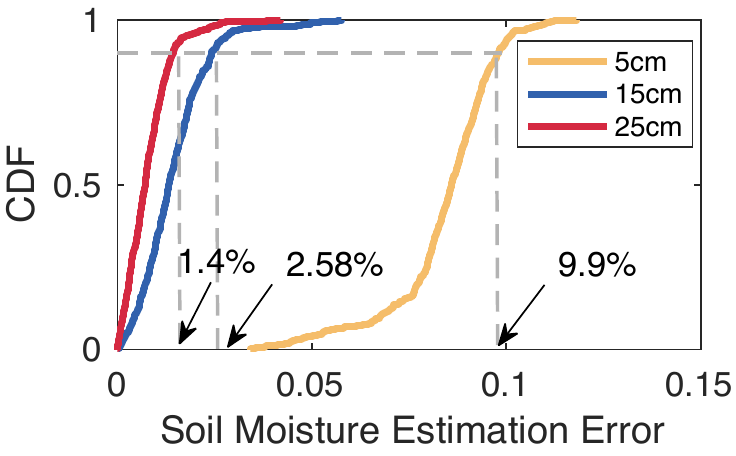}
\caption{Sensing accuracy \textit{vs.} deployment depth.}
\label{fig:depth_cdf}
\end{minipage}
\hspace{2mm}
\begin{minipage}[!t]{0.34\linewidth}
\setlength{\abovecaptionskip}{1mm}
\setlength{\belowcaptionskip}{-5mm}
\centering
\includegraphics[width=0.95\linewidth]{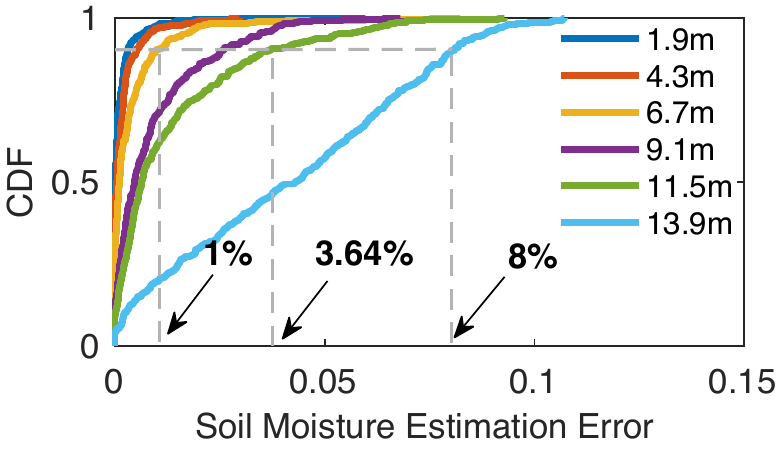}
\caption{Sensing accuracy \textit{vs.} working distance.}
\label{fig:acc_distance}
\end{minipage}
\hspace{2mm}
\begin{minipage}[!t]{0.29\linewidth}
\setlength{\abovecaptionskip}{1mm}
\setlength{\belowcaptionskip}{-5mm}
\centering
\includegraphics[width=0.95\linewidth]{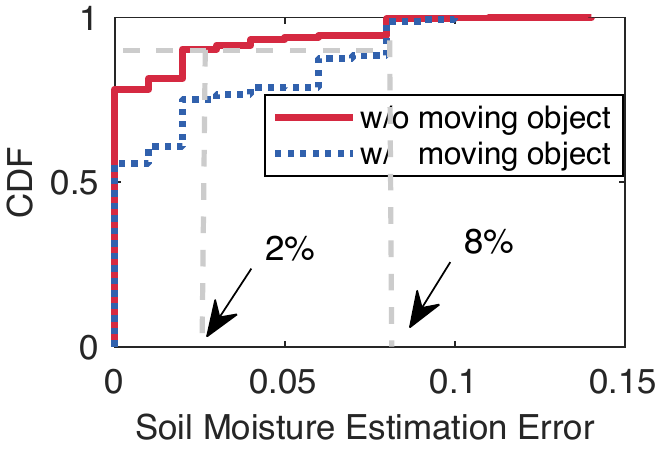}
\caption{Sensing accuracy \textit{vs.} environmental noise.}
\label{fig:interference}
\end{minipage}
\end{figure*}

Fig.~\ref{fig:acc_distance} shows the CDF of moisture estimation errors under different Tx-tag distances. We have two observations. First, when the Tx-to-Tag distance is less than 6.7~m, \systemname achieves a high accuracy with the 90th percentile moisture estimation errors of 1.78\%. Second, the sensing accuracy decreases with the increase of distance. Overall, \systemname still achieves a high moisture sensing accuracy of 8\% even when the working range is 13.9~m. While, the state-of-the-art wireless moisture sensing system GreenTag~\cite{wang2020Soil} only works within the range of 2~m.

\subsubsection{Impact of the environment variations}\label{sec:evaluation:motion}
In real application scenarios, environment variations, such as the movement of human body, would cause signal variations, which may result in estimation errors. Thus, we evaluate the impact of environment variations on the sensing accuracy.
To create a dynamic environment, we ask one person to move nearby the experiment deployment. The distance between the moving person and transceivers is within 0.2$\sim$1~m, and the person continuously walks around during the data collection period. For each soil moisture sample, we collect 2500 packets with and without human motion, and then calculate the estimation errors. Fig.~\ref{fig:interference} shows the CDF of soil moisture estimation errors with and without human motion. Compared with the static environment, the $90th$ estimation error increases from 2\% to 8\% when there are human motions. Because the human movement would introduce some interference signals, making the signal features of some soil moisture may overlap and thus leading to estimation errors. Overall, \systemname still achieves an 8\% moisture sensing accuracy even in a dynamic environment.

\subsubsection{Impact of different types of environments}\label{sec:evaluation:environment}

To analyze the impact of different multipath environments on the sensing accuracy, we also conduct experiments in five kinds of environments since different environments have different multipath distribution.
The five environments are: $(i)$ the anechoic chamber which is a static environment without multipath; $(ii)$ the open room which is a static environment with less multipath; $(iii)$ the home courtyard with less multipath and human motion; $(iv)$ the open farmland with static environment where vegetables are grown; $(v)$ the office room with more multipath and static environment. In the five environments, the soil used in all experiments was packed in a large plastic box with a capacity of 45~L as shown in Fig.~\ref{fig:loam_soil}b. The tag is buried at a depth of 25~cm. The experiments are carried out in May and the weather is sunny with $27^{\circ}C$$\sim$$30^{\circ}C$ air temperature. In the five environments, the Tx-Tag distance is set to 3~m.

\begin{figure*}[t]
\setlength{\abovecaptionskip}{0mm} 
\setlength{\belowcaptionskip}{1mm}
\centering
\begin{minipage}[t]{0.28\linewidth}
\setlength{\abovecaptionskip}{0mm} 
\setlength{\belowcaptionskip}{1mm}
\centering
\includegraphics[width=0.95\linewidth]{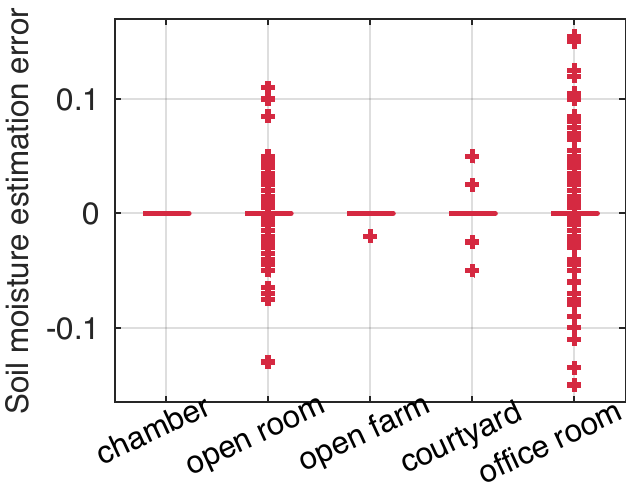}
\caption{The soil moisture estimation error in 5 environments.}
\label{fig:5_environment}
\end{minipage}
\hspace{25mm}
\begin{minipage}[t]{0.35\linewidth}
\setlength{\abovecaptionskip}{0mm} 
\setlength{\belowcaptionskip}{1mm}
\centering
\includegraphics[width=0.97\linewidth]{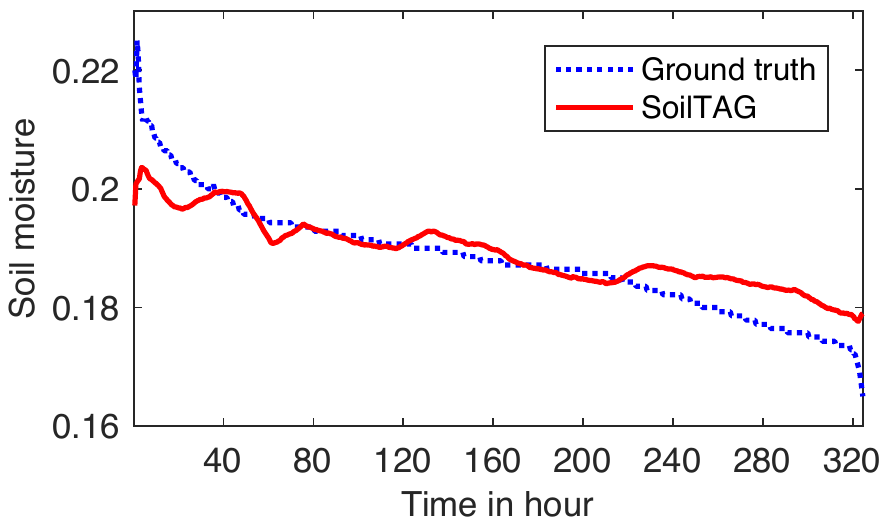}
\caption{Continuously soil moisture monitoring in a soil box.}
\label{fig:case1}
\end{minipage}
\end{figure*}

Fig.~\ref{fig:5_environment} shows the boxplot of the soil moisture estimation error in the five environments. The outlier point is lower than $10th$ percentile error value or higher than $90th$ percentile error value. We can see that the $10th$$\sim$$90th$ percentile soil moisture estimation error of all environments are 0\%, which further demonstrates the high accuracy of our system on soil moisture sensing. In addition, we also find that the number of error outliers gets more with the environment gets more complex. This is because the multipath and human motion would distort some relative frequency response features. As we expected, the chamber environment has no error outlier and the office room has the most error outliers. The open farm and courtyard has little error outliers. This result implies that our system is suitable for the common soil moisture application scenarios (i.e. the farm field and courtyard).

\subsection{Case Study : Continuously soil moisture monitoring application}\label{sec:case}
Finally, we conduct a case study to monitoring soil moisture changing continuously. We configure the transceiver with directional antennas respectively and set Tx-Tag and Rx-Tag distance as 1~m respectively. We bury the chipless Wi-Fi tag into the soil at the depth of 15~cm to detect soil moisture. The tag has a resin material coat to prevent it from shorting circuit by the soil. The tag used in this experiment is shown in Fig.~\ref{fig:tag}. The system automatically collects data every half an hour. In order to measure the ground truth, we manually measured the soil moisture every half an hour using Delta-T Wet-2 sensor.

Fig.~\ref{fig:case1} plots the soil moisture value in two weeks. The blue line is the ground truth and the red line is the estimated value of our system.
We can see that the two lines show the same changing trend and the red line is basically consistent with the blue line. We also find that there are some points that deviate from the ground truth. The reason is that there is human motion during the data collection period. In addition, we can see that the ground truth soil moisture decreased by 5.5\% from 22.5\% to 17\% . 
However, it is difficult for most cheap soil sensors to stably detect such small changes in soil moisture. Actually, for some plants a slight soil moisture changing would affect its growth. Such application requires a system that is able to detect small changes in soil moisture, to tell the farmer irrigate the plant timely. This result demonstrate our system is capable of continuously monitoring soil moisture changing in real-world farm.

\section{Conclusion}
This paper presented {\systemname}, a battery-free Wi-Fi tag based high accuracy, low-cost and long range soil moisture sensing system. To measure the soil moisture, our idea is converting changes in soil moisture levels into the frequency response of the tag's resonator. To achieve high accuracy, we optimize key resonator parameters to increase the frequency response sensitivity for sensing even small soil moisture changes. To improve the working range, we design patch antenna array for the tag and employ beamforming and beam nulling at the transceiver side. Experimental results show that \systemname achieves a 2\% high accuracy when the working range is less than 6 m, and when the working range is up to 10 m, it achieves 3.64\% accuracy. We believe \systemname has promising potential for improving the productivity of farm field and our exploration about passive sensor design can enrich other passive sensing applications.
\bibliographystyle{plain}
\bibliography{reference}
\end{document}